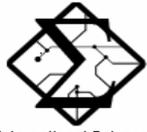

# Towards Design and Implementation of a Language Technology based Information Processor for PDM Systems

[1,2]Zeeshan Ahmed; [2]Saman Majeed, [2]Thomas Dandekar

[1]Vienna University of Technology Austria & [2]University of Wuerzburg Germany

Authors' E-mails: zeeshan.ahmed@htomail.de, saman.majeed@uni-wuerzburg.de, dandekar@biozentrum.uni-wuerzburg.de

Url: www.uni-wuerzburg.de, www.tuwien.ac.at and www.zeeshanahmed.info

*Abstract*—**Product Data Management (PDM)** aims to provide 'Systems' contributing in industries by electronically maintaining organizational data, improving data repository system, facilitating with easy access to CAD and providing additional information engineering and management modules to access, store, integrate, secure, recover and manage information. Targeting one of the unresolved issues i.e., provision of natural language based processor for the implementation of an intelligent record search mechanism, an approach is proposed and discussed in detail in this manuscript. Designing an intelligent application capable of reading and analyzing user's structured and unstructured natural language based text requests and then extracting desired concrete and optimized results from knowledge base is still a challenging task for the designers because it is still very difficult to completely extract Meta data out of raw data. Residing within the limited scope of current research and development; we present an approach capable of reading user's natural language based input text, understanding the semantic and extracting results from repositories. To evaluate the effectiveness of implemented prototyped version of proposed approach, it is compared with some existing PDM Systems, in the end the discussion is concluded with an abstract presentation of resultant comparison amongst implemented prototype and some existing PDM Systems.

## 1. Introduction

Product Data Management Systems are heavily contributing in industries by electronically maintaining organizational data, improving data repository system and providing information management module controls to access, store, integrate, secure, recover and manage information. Moreover PDM Systems are also capable of establishing basic networked computer environment, implementing interface module to support user queries, providing information structure management to produce exact structured information and system administration to set up administration and configuration of the system [1]. Where PDM Systems are heavily benefiting the industries there PDM community is also facing some serious unresolved issues i.e. *Successful Platform Independent System Implementation, Easy and friendly PDM System deployment and Reinstallation mechanism, Static and Unfriendly Human Machine Interface (HMI), Static and Unintelligent Search, Insecurity and non standardized Framework*.

Targeting these mentioned problems, many approaches and solutions including *Metaphase (SDRC), SherpaWorks (Inso), Enovia (IBM), CMS (WTC), Windchill (PTC), and Smarteam (Smart Solutions)* [2] etc have been proposed, but these issues are still unresolved. Targeting these alive unresolved issues we have also proposed a semantic oriented information engineering based solution i.e., *Intelligent Semantic Oriented Agent based Search* (I-SOAS) towards PDM community. As shown in Fig. 1, the main concept of I-SOAS is divided into four sequential iterative components i.e. Intelligent

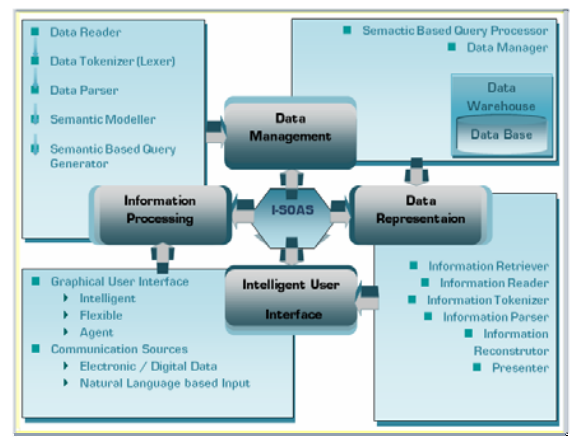

Fig.1. I-SOAS Conceptual Model [3]

User Interface (IUI), Information Processing (IP), Data Management (DM) and Data Representation (DR). IUI is proposed to design intelligent graphical user interface for efficient system user communication, IP is proposed to process and model user's unstructured and structured inputted request by reading, lexing, parsing, and semantic modeling, DM is proposed to manage user requests and system performance based information and DR is proposed to represent system's output in user's understandable format [4]. To implement the I-SOAS conceptual model in to a real time software application, we have designed an implementation architecture of I-SOAS consisting of four main modules i.e. *Graphical Interface, Processing and Modeling, Repository and Knowledge Base* and three communication layers i.e.

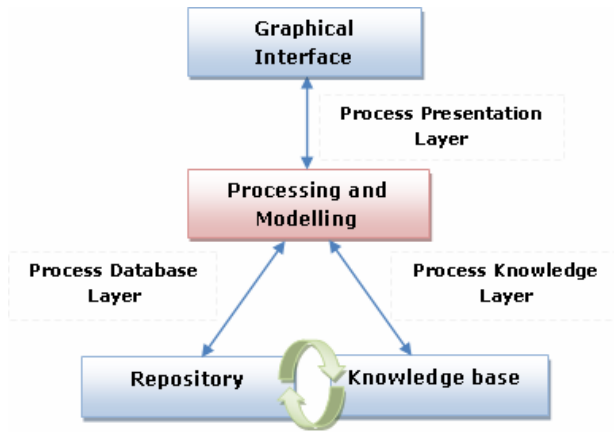

Fig.2. I-SOAS Implementation Architecture

Process Presentation Layer, Process Database Layer, and Process Knowledge Layer [5] (please See Fig 2). The graphical Interface is a flexible and intelligent user system communication interface of the application, Repository is the database to store and manage data and Knowledge base is to capture, manage, improve and deliver knowledge. Processing and Modeling is to read, organized, tokenize, parse, semantically evaluate and generate semantic based queries to extract desired results from Repository and Knowledge base. Moreover three communication layers are designed to transfer data between above mentioned four major modules of I-SOAS Implementation Architecture.

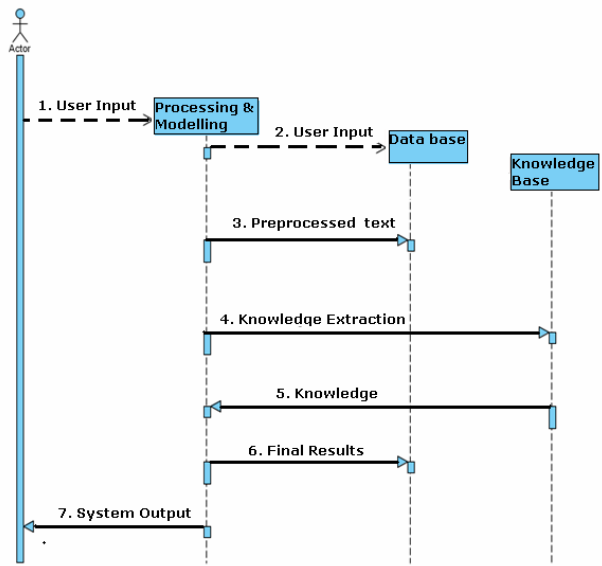

Fig.3. I-SOAS System Sequence Design

PPL is responsible for data transfer between Graphical Interface and Processing and Modeling, PDL is responsible for data transfer between Repository and Processing and Modeling, whereas PKL is responsible for data transfer between Knowledge base and Processing and Modeling.

The designed system sequence as shown in Fig 3, works in a way that the graphical user interface is used to take the input from user and forward to the Processing and Modeling component which first saves the user input in to Database then processes to extract Knowledge from Knowledge base and in the end restores the final results into database and provides resultant output to the user via graphical user interface.

The overall development of each component of I-SOAS is performed in Java programming language [6] but for the development of Graphical Interface one of the RIA technology [24] called Flex [25] is used. MySQL is used for the construction of Data Storage [26]. Grammar is written using Antlr [27] and for the development of Knowledge Base and the construction of ontologies semantic web technologies XML [7], RDF [8] and OWL [9] are used (please See Fig. 4).

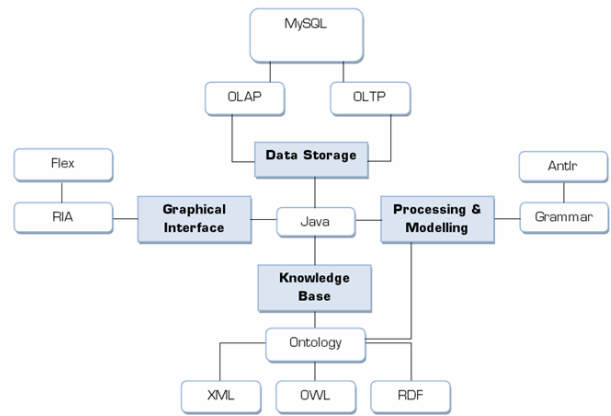

Fig.4. Involved Technologies

In this manuscript we are not going into the details of any module I-SOAS except Information Processing & Modeling module. Later in this manuscript, at first we briefly discuss the most relevant literature to our research work as the part of state of art. Then we present Processing & Modeling in detail and conclude the discussion with the presentation of implemented prototype.

## 2. State of the Art – Language Technology

Relevant literature as the part of related research work is collected from several sources like proceedings of conferences, scientific journals and project websites etc. We used web based search engines i.e., Google [10] and Wikipedia [11] for querying to extract and review the relevant information.

In the domain of information engineering and semantic modeling many methodologies, propositions and approaches have been introduced by many researchers but we are only considering the Language Technology [13] based work as the most relevant literature to our research work presented in this paper.

Language Technology is a linguistic based field of computer science, also called as Human Language Technology or Natural Language Processing [12], about to make machine capable to read, listen, understand and analyze human (natural) language. The main objective of Languages Technology is to teach machines, how to communicate and help humans by communicating (listening and speaking) with them [13]. Although existing Language Technology based desktop, web and semantic web applications are not able to completely read, understand and analyze human (natural) language based instructions. But still the community of Language Technology is struggling in producing such intelligent systems which will be based on natural language interfaces capable of communicating with human in natural language by listening, understanding and analyzing the context and semantic.

There are many ongoing language technology based research projects contributing in cross lingual Information, knowledge management, multilingual document production and multilingual natural communication like *TWENTYONE [14], MULINEX [15], MIETTA [16], OLIVE [17], PARADIME [18], Whiteboard [19], GETESS [20], TG/2 [21] and TEMSIS [22]* etc.

*TWENTYONE:* A European Commission sponsored project for disclosure and dissemination of documents on sustainable development for environmental organizations by locating and automatic translation of information [14].

*MULINEX*: A European Commission sponsored project for multilingual indexing, navigation and editing extensions for efficient use of multilingual online information by providing the combinations of the newest Information Retrieval Technology with advanced Language Technology to improve search and navigation in the WWW [15].

*MIETTA*: Multilingual Tourist Information on the World Wide Web is a European Commission sponsored project for providing flexible cross-lingual access to tourist information in the web, combination of class-based and free text search, Simultaneous access to heterogeneous data sources, presentation of information in different languages through machine translation and multilingual generation and Advanced localization and web-technology for simplified maintenance [16].

*OLIVE*: Retrieval of video material based on speech-recognition is a European Commission sponsored project for providing Video-retrieval on the basis of indices constructed from transcribed speech, three step retrieval process from textual index terms, cross-lingual retrieval on the basis of document translation and query translation, additional background material integrated for improved speech recognition and for further disclosure of video contents [17].

*PARADIME*: Intelligent Extraction of Information from On-line Documents is a European Commission sponsored project for providing intelligent retrieval of information from German documents, management of vast sources of linguistic knowledge, new functions with machine learning processes and integration of graphic visualization techniques, server architecture and access to the Internet [18].

*Whiteboard*: Multilevel Annotation for Dynamic Free Text Processing is BMB+F sponsored project for designing, implementing, investigating and evaluating a new system architecture that facilitates the combination of different language technologies for a range of practical applications [19].

*GETESS*: German Text Exploitation and Search System is BMB+F sponsored project for the development of an intelligent work tool for researching information [20].

*TG/2*: Practical generation of natural language text, TG/2 stands for a new generation of template based generators and designed to organize a classical production system, separating the generation rules from their interpreter. TG/2 can provide solutions for limited sublanguages that are tuned towards the domain, can quickly be accommodated to new tasks, can be integrated smoothly with 'deep' generation processes, can reuse generated substrings for alternative formulations and can be parameterized to produce the preferred formulations[21].

*TEMSIS*: Transnational Environmental Management Support & Information System is a European Commission sponsored project for increasing availability of up-to-date data in the information age is of limited use without adequate presentation. TEMSIS can provide On-line access to current measurements, can select Information according to the user's requirements, can generate environmental reports in the user's language (German, French), can build data and parameter dependent structure, can combine extemporaneous and fixed text components and can represent, comprise and summaries collected data [22].

*COSMA*: Automated Appointment Scheduling by E-Mail provides concepts of automatic planned and manage appointments using software agents to save time and money savings during the arrangement of business appointments by communicating with a number of partners via e-mail, distributing client/server solution in the Internet, robusting dialogue control with intelligent failure handling and Shallow-Parsing techniques for the analysis of the German language [23].

Language Technology seems very important field with respect to our research goals and has much relevancy to our research and development work. As Language Technology aims to teach machines, how to communicate and help humans by communicating (listening and speaking) with them which is one of the key interests and very important goal of our research and development work. Moreover the followed or proposed methodologies by the Language Technology community in fulfilling their goals of research can also be very helpful in our research and development work.

### 3. I-SOAS Information Processing & Modeling

Designing an intelligent application capable of reading and analyzing user's structured and unstructured text based requests and then extracting desired concrete and optimized results from knowledge base is still a challenging task for the designers because still it is very difficult to completely extract Meta data out of data. I-SOAS Information Processing & Modeling module is a semantic based system proposed and designed to read, organize, tokenize, parse, semantically evaluate, model and process information to extract desired results from data repository and knowledge base.

*3.1. Design Requirements*
Information Processing and Modeling is designed keeping five major requirements in mind *i.e. the designed PM module must be capable of*
1. Handling user inputted request by analyzing the inputted instructions
2. Understanding the context and semantic of user's requested input by lexing, parsing, processing and modeling instructions
3. Storing and managing the user's requested input and final output in database
4. Extracting user's requested knowledge and Producing system output in user understandable format

*3.2. Work Flow*
To take advantage and organize the implementation process of proposed and designed Information Processing and Modeling in a real time software application, we have designed an internal work flow model (please see Fig. 5). Information Processing and Modeling internal work flow model starts with the input text, which then moves to I-SOAS Lexer after storing into the I-SOAS Database. I-SOAS Lexer first tokenizes the input text into possible number of token, evaluates the context with respect to the used grammar of the input language and then forward to I-SOAS Parser.

    I-SOAS Parser considers all the statements or instructions as the combination of all tokens produced by I-SOAS Lexer for semantic evaluation and forward resultant semantically evaluated information to I-SOAS Modeler. Then, I-SOAS Modeler produces semantic based models using syntactically and semantically evaluated information and forward resultant modeled information to I-SOAS Resolver after storing into database. I-SOAS Resolver takes semantic models as input and then resolves these models by producing relationships between modeled statements. Then finally resolved relation based information is stored in to database and can be treated as the final output of I-SOAS Information Processing.

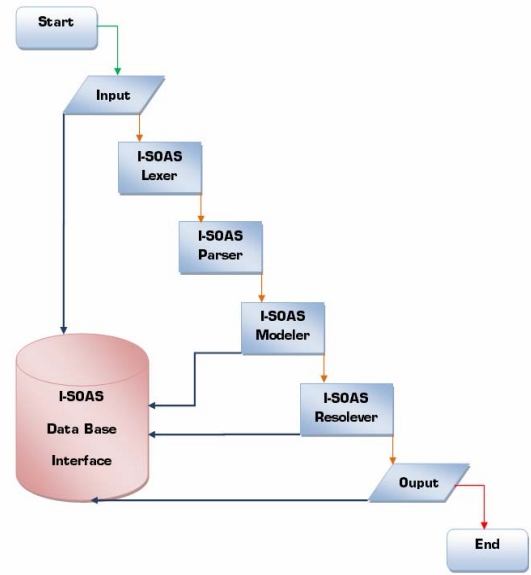

Fig 5. Work Flow

*3.3. System Sequence*
We implement proposed and designed Information Processing and Modeling in a way that not only it will perform its individual jobs but can also work in an integrated form by sending and receiving information to the other connected modules of the I-SOAS implementation architecture via layers (please See Fig. 2).

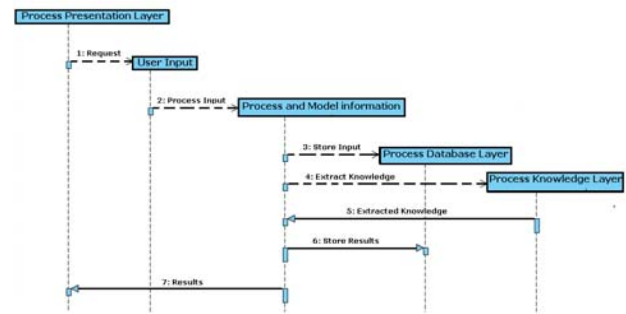

Fig 6. System Sequence Design

The information Processing and Modeling system sequence diagram consists of five main stepped activities i.e. *Process Presentation Layer, User Input, Process and Model Information, Process Database Layer and Process Knowledge Layer* (see Fig. 6). These five stepped activities are designed to perform certain jobs. The job of Process Presentation Layer is to bring user requested input from Graphical Interface to the User Input component of Information Processing and Modeling. Then User Input component will forward the user requested inputted instructions to Process and Model Information. Process and Model Information component will first store the information in to Repository via Process Database Layer, then will process the information by lexing, parsing and semantically modeling, and then extract needed knowledge from Knowledge Base via Process Knowledge Layer. In the end Process and Model Information will first save the information in I-SOAS Database via Process Database Layer (PDL), and then return the final system output to I-SOAS Graphical Interface via Process Presentation Layer (PPL).

*3.4. Ontological View*

Following the concepts of Antlr [27] for language grammar writing, we have written a basic English language based grammar (Introduction - see Appendix) which is then implemented into an ontological view using the concepts of semantic web [28] and ontology [29]. As shown in Fig 7 the overall ontology structure consists of one main Class ISOAS, then three example subclasses of ISOAS are i.e. *A, B and C*.

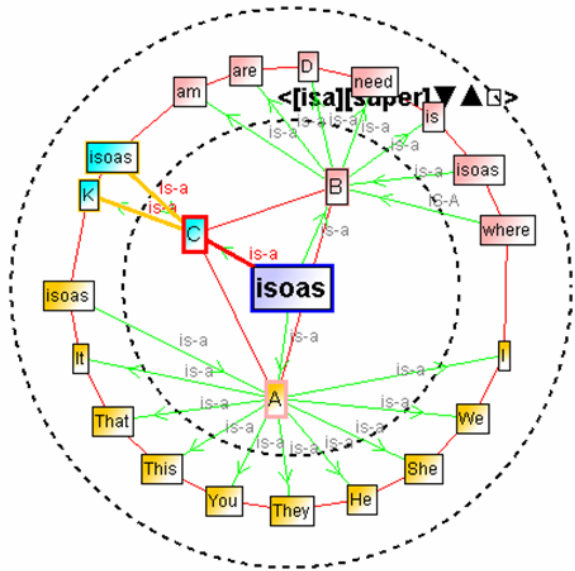

Fig.7. I-SOAS Ontological View – "isa" equality relation (see Fig. 8)

All three subclasses contain their further subclasses and the relationships of subclass with each other. Class A contains eight sub classes i.e. *I, We, He, She, You, It, This, That' and They*. Class B contains seven sub classes i.e. *is, are, am, need, want where and D*. D is another subclass containing two more subclasses i.e. *looking and Searching*, and looking have three more subclasses i.e., *for , between and equal*. Class C contains only one sub class K. To give for this simple example more details (see Fig. 8).

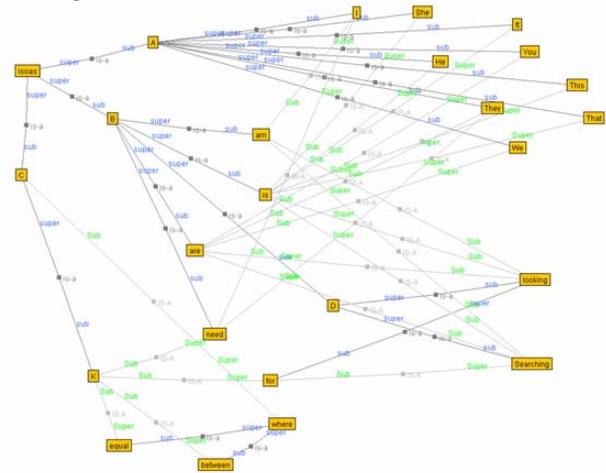

Fig.8. Class Relationships (introductory example)

Class A has direct relationship with its subclasses,
- "I" has the relationship with "am"
- "He" has the relationship with "is"
- "She" has the relationship with "is"
- "It" has the relationship with "is"
- "That" has the relationship with "is"
- "This" has the relationship with "is"
- "They" has the relationship with "are"
- "We" has the relationship with "are"

Class B has direct relationship with its subclasses,
- "is" has the relationship with "looking" and "searching"
- "am" has the relationship with "looking" and "searching"
- "are" has the relationship with "looking" and "searching"
- "need" has the relationship with "K"
- "need" has the relationship with "Where"
- "want" has the relationship with "Where"
- "want" has the relationship with "K"
- "where" has the relationship with "between"
- "where" has the relationship with "equal"

Class C has direct relationship with its subclasses "K", and "K" has relationship with "need", "want" and "for"

## 4. I-SOAS Prototype

As this research paper is about an ongoing process in research work, following the constructed implementation designs and meeting the design requirements, a prototype version of I-SOAS is developed (see Fig 9).
The currently available version of I-SOAS Desktop Application is capable of
- Running as a stable application.
- Taking input from user in text format.

- Applying Lexer to produce tokens.
- Applying Parser to evaluate the semantic.
- Providing dynamic database management system to storing user input, processed data and system output

Producing different kind of visualizations of stored user based and system processed information in database.

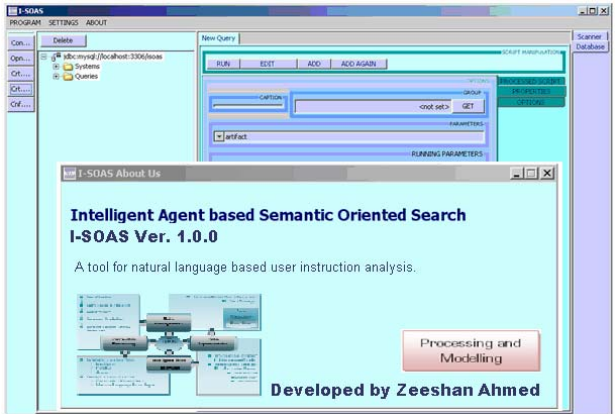

Fig.9. I-SOAS Prototype

## 5. Comparison

A comparison is performed amongst currently available version of I-SOAS and two different PDM systems i.e., *Windchill [30] and CDB [31]*, resultant information is presented in table 1.

Table 1. Comparison

| No | Jobs | Wind-chill | CDB | I-SOAS |
|---|---|---|---|---|
| 1 | Static search using fixed options | Yes | Yes | Yes |
| 2 | Limited search | Yes | Yes | Yes |
| 3 | Natural language based search | No | No | Yes |
| 4 | Creating dynamic database to store user search based information | No | No | Yes |
| 5 | Connect and disconnect different databases to search | No | No | Yes |
| 6 | Manipulating different database to search and store information | No | No | Yes |
| 7 | Editor to build manual search queries to extract stored results | No | No | Yes |
| 8 | Processing and modeling unstructured and structured data | No | No | Yes |
| 9 | Create, Edit, Save and Run search query | No | No | Yes |
| 10 | Integrate search queries | No | No | Yes |
| 11 | SQL Query Mode of search | No | No | Yes |
| 12 | Provide searched results back to the user | Yes | Yes | Yes |

As described in table 1, the currently available prototype version of I-SOAS has several many advantages over Windchill and CDB with respect to the Search Mode. The three compared PDM products have limited fixed options based static search. While I-SOAS compensates for this feature by providing an intelligent search based on defined natural language processing.

## 6. Conclusion

In this research paper we have briefly described Product Data Management and some major existing challenges, then, continuing the presentation of research, briefly described the conceptual and implementation architecture of our own proposed solution towards PDM challenges i.e. I-SOAS. In this research paper we mainly focused on the design implementation of an important module of I-SOAS implementation architecture i.e. Information Processing and Modeling. Going into the details of this module we have presented the information about the theme, design requirements, designs and technologies involved in the development.

## 7. Future Recommendations

As this is an in process and on going research, in future, we are aiming to present a web based prototype version of I-SOAS capable of not only providing all capabilities for natural language based search available in existing implemented desktop prototype of I-SOAS but with also some more functionalities for better system user communication and intelligent search.

## 8. References

[1] A. Zeeshan, D. Gerhard: "Contributions of PDM Systems in Organizational Technical Data Management", In First IEEE International Conference On Computer, Control & Communication, Karachi Pakistan , November 2007

[2] M.Y. Huang, Y.J. Lin, Hu Xu, "A framework for web-based product data management using J2EE", in *The International Journal of Advanced Manufacturing Technology, Springer London, ISSN 0268-3768 (Print) 1433-3015 (Online), Volume 24,pp 847-852, Numbers 11-12,* December 2004

[3]A. Zeeshan, D. Gerhard: "Intelligent Graphical User Interface, Information Processing and Data Management based Solution towards Intelligent Data Manipulation and Representation", In *4th Virtual Conference of the EU-funded FP6 I*PROMS Network of Excellence on Innovative Production Machines and Systems*, Cardiff England, 2008

[4] A. Zeeshan, D. Gerhard: "Semantic Oriented Agent based Approach towards Engineering Data Management, Web Information Retrieval and User System Communication Problems", *In the proceedings of 3rd International Conference for Internet Technology and Secured Transactions*, Dublin Ireland, June 2008

[5] A. Zeeshan, D. Gerhard: "Design Implementation of Semantic Oriented Agent and Knowledge based approach for Intelligent Human Machine Data Manipulation ", In 4th Virtual International Conference on Innovative Production Machines and Systems, Cardiff UK, 2008

[6] Java, reviewed 25 September 2008,<http://www.java.com/en/> , 2008

[7] The W3C Extensible Markup Language (XML), viewed February 2007, <http://www.w3.org/XML>, 2007

[8] Klaus Tochtermann and Herman Maurer, "Semantic technologies – An Introduction", in *Semantic Technologies Showcase the Austrian Situation, pp. 15-20*, 2006

[9] OWL Web Ontology Language, viewed February 2007, <http://www.w3.org/TR/owl-features>, 2007

[10] Google, viewed December 2008, <http://www.google.com>, 2008

[11] Wikipedia, viewed December 2008, < www.wikipedia.org >, 2008

[12] Language technology, Reviewed 03 September 2008, <http://en.wikipedia.org/wiki/Human_language_technology>

[13] Hans Uszkoreit, DFKI-LT - What is Language Technology?, Reviewed 03 September 2008, <http://www.dfki.de/lt/lt-general.php>

[14] TWENTYONE: Disclosure and dissemination of documents on sustainable development, Reviewed 03 September 2008, <http://www.dfki.de/pas/f2w.cgi?ltc/twentyone-e>

[15] MULINEX: Multilingual Indexing,Navigation and Editing Extensions for the WWW, Reviewed 03 September 2008, <http://www.dfki.de/pas/f2w.cgi?ltc/mulinex-e>

[16] MIETTA: Multilingual Tourist Information on the World Wide Web, Reviewed 03 September 2008, <http://www.dfki.de/pas/f2w.cgi?ltc/mietta-e>

[17] OLIVE: Play it again, Sam! Retrieval of video material based on speech-recognition, Reviewed 03 September 2008, <http://www.dfki.de/pas/f2w.cgi?ltc/olive-e>

[18] PARADIME: Intelligent Extraction of Information from On-line Documents, Reviewed 03 September 2008, <http://www.dfki.de/pas/f2w.cgi?ltc/paradime-e>

[19] Whiteboard: Multilevel Annotation for Dynamic Free Text Processing, Reviewed 03 September 2008, <http://www.dfki.de/pas/f2w.cgi?ltc/whiteboard-e >

[20] GETESS: German Text Exploitation and Search System, Reviewed 03 September 2008, <http://www.dfki.de/pas/f2w.cgi?ltc/getess-e>

[21] TG/2: Practical generation of natural language text, Reviewed 03 September 2008, <http://www.dfki.de/pas/f2w.cgi?lts/tg2-e>

[22] TEMSIS: Transnational Environmental Management Support & Information System, Reviewed 03 September 2008, <http://www.dfki.de/pas/f2w.cgi?ltc/temsis-e>

[23] COSMA: Automated Appointment Scheduling by E-Mail, Reviewed 03 September 2008, <http://www.dfki.de/pas/f2w.cgi?lts/cosma-e>

[24] Rich Internet application, Last reviewer 15-02-2009, <http://en.wikipedia.org/wiki/Rich_Internet_application>

[25] Adobe Flex 3, Last reviewer 15-02-2009, <http://www.adobe.com/products/flex/>

[26] MySQL, Last reviewer 15-02-2009, <http://www.mysql.com/>

[27] Antlr, last reviewed 15-02-2010, <http://www.antlr.org/>

[28] Semantic Web, last reviewed 15-02-2010, <http://en.wikipedia.org/wiki/Semantic_Web>

[29] Ontology, last reviewed 15-02-2010, <http://en.wikipedia.org/wiki/Ontology>

[30] Windchill, last reviewed 15-02-2010, <http://www.ptc.com/products/windchill/>

[31] CONTACT Software presents preview of CIM DATABASE 2.9.8, last reviewed 15-02-2010, <http://www.contact.de/news/archive/news2009/P08-01-09UK>

## 9. Appendix

```
/*START*/
Digits:     ( '0'..'9') ;
NUMBER :          (DIGIT)+ ;
A:         'I'|'We'|'They'|'He'|'She'|'You'|'It'|'This'|'That';

B:         'need'|'want'|'look for'|'look about'|'search'|'ask for'|'seek for'|'needs'|'wants'|'looks for'|'looks about'|'seraches'|'asks for'|'seeks for'|'am looking for'|'am searching'|'am asking for'|'am seeking'|'am in search of'|'are looking for'|'are searching'|'are asking for'|'are seeking'|'are in search of'|'is looking for'|'is searching'|'is asking for'|'is seeking'|'is in search of'|'define'|'quest'|'questing'|'identify'|'scratch around'|'root about'|'find'|'after';

C:
           ('0'..'9'|'a'..'z'|'A'..'Z'|'PDM'|'CAD'|'document'|'printer'|'presentation'|'application'|'contract'|'office'|'section'|'quarter'|'airport'|'boulevard'|'street'|'country'|'sity'|'town'|'shop'|'busstop'|'hotel'|'hostel'|'theater'|'cinema'|'movies'|'picture'|'film'|'song'|'singer'|'music'|'lyrics'|'radio'|'group'|'game'|'news'|'job'|'train station'|'torrents'|'subtitles'|'gifts'|'clothes'|'shoes'|'dress'|'clothing'|'banks'|'weather'|'books'|'magazines'|'newspaper'|'publications'|'articles'|'events'|'concerts');

W:         'where';

Bt:        'between';

Eq:        'equal to'|'with'|'less than'|'greater than'|'less than and equal to'|'greater than and equal to '|'='|'>'|'<'|'<='|'>=';

And:       'and';

WS:        (' '|'\r'|'\t'|'\u000C'|'\n')  {$channel=HIDDEN;}   ;

/*Direct Search Rules*/

astmt      : A;

bstmt      : B;

cstmt      : C;

stmt1      : A B C;

stmt2      : B C;

/*Conditional Search Rules*/

condbt     : A B C W Bt C And C ;

condeq     : A B C Eq C;

condweq    : A B C W Eq ;

condeqbt   :A B C W Eq Bt C And C ;

/*END*/
```

## 10. Acknowledgement
We thanks all those who have been involved with the research and development of I-SOAS over the years.

## 11. Author's Biography

Zeeshan Ahmed, the main author of this paper, presently a Research Scientist at University of Wuerzburg. He has on record more than 12 years of University Education and more than 7 years of Professional Experience of working within different multinational organizations in the field of Computer Science with emphasis on software engineering of artificially intelligent applications and product line architectures.

Coauthor; Saman Majeed, is a promotion admission student at University of Wuerzburg.

Coauthor; Thomas Dandekar is a research supervisor, Chari Department of Bioinformatics, University of Wuerzburg.